\documentclass[twocolumn,prb,showpacs,superscriptaddress,preprintnumbers,amsmath,amssymb]{revtex4}

\usepackage{graphicx} 
\usepackage{dcolumn} 
\usepackage{bm} 

\usepackage{tabularx}
\newcolumntype{L}[1]{>{\raggedright\arraybackslash}p{#1}} 
\newcolumntype{C}[1]{>{\centering\arraybackslash}p{#1}} 
\newcolumntype{R}[1]{>{\raggedleft\arraybackslash}p{#1}} 

\begin{document}

\title{Crystal-field ground state of the orthorhombic Kondo insulator CeRu$_2$Al$_{10}$}

\author{F.~Strigari}
  \affiliation{Institute of Physics II, University of Cologne, Z{\"u}lpicher Stra{\ss}e 77, 50937 Cologne, Germany}
\author{T.~Willers}
  \affiliation{Institute of Physics II, University of Cologne, Z{\"u}lpicher Stra{\ss}e 77, 50937 Cologne, Germany}
\author{Y.~Muro}
  \affiliation{Department of Quantum Matter, AdSM, Hiroshima University, Higashi-Hiroshima 739-8530, Japan}
\author{K.~Yutani}
	\affiliation{Department of Quantum Matter, AdSM, Hiroshima University, Higashi-Hiroshima 739-8530, Japan}
\author{T.~Takabatake}
	\affiliation{Department of Quantum Matter, AdSM, Hiroshima University, Higashi-Hiroshima 739-8530, Japan}
	\affiliation{Institute for Advanced Materials Research, Hiroshima University, Higashi-Hiroshima 739-8530, Japan}
\author{Z.~Hu}
  \affiliation{Max Planck Institute for Chemical Physics of Solids, N{\"o}thnizer Stra{\ss}e 40, 01187 Dresden, Germany}
\author{Y.-Y. Chin}
  \affiliation{Max Planck Institute for Chemical Physics of Solids, N{\"o}thnizer Stra{\ss}e 40, 01187 Dresden, Germany}
\author{S.~Agrestini}
  \affiliation{Max Planck Institute for Chemical Physics of Solids, N{\"o}thnizer Stra{\ss}e 40, 01187 Dresden, Germany}
\author{H.-J.~Lin}
  \affiliation{National Synchrotron Radiation Research Center (NSRRC), 101 Hsin-Ann Road, Hsinchu 30077, Taiwan}
\author{C.~T.~Chen}
  \affiliation{National Synchrotron Radiation Research Center (NSRRC), 101 Hsin-Ann Road, Hsinchu 30077, Taiwan}
\author{A.~Tanaka}
  \affiliation{Department of Quantum Matter, AdSM, Hiroshima University, Higashi-Hiroshima 739-8530, Japan}
\author{M.~W.~Haverkort}
  \affiliation{Max Planck Institute for Solid State Research, Stuttgart, Germany}
\author{L.~H.~Tjeng}
 \affiliation  {Max Planck Institute for Chemical Physics of Solids, N{\"o}thnizer Stra{\ss}e 40, 01187 Dresden, Germany}
\author{A.~Severing}
  \affiliation{Institute of Physics II, University of Cologne, Z{\"u}lpicher Stra{\ss}e 77, 50937 Cologne, Germany}

\date{\today}

\begin{abstract}
We have succeeded in establishing the crystal-field ground state of CeRu$_2$Al$_{10}$, an orthorhombic intermetallic compound recently identified as a Kondo insulator. Using polarization dependent soft x-ray absorption spectroscopy at the Ce $M_{4,5}$ edges, together with input from inelastic neutron and magnetic susceptibility experiments, we were able to determine unambiguously the orbital occupation of the $4f$ shell and to explain quantitatively both the measured magnetic moment along the easy $a$ axis and the small ordered moment along the $c$-axis. The results provide not only a platform for a realistic modeling of the spin and charge gap of CeRu$_2$Al$_{10}$, but demonstrate also the potential of soft x-ray absorption spectroscopy to obtain information not easily accessible by neutron techniques for the study of Kondo insulators in general.
\end{abstract}


\maketitle

CeRu$_2$Al$_{10}$ is a fairly new synthesized orthorhombic intermetallic compound \cite{Thiede1998,Tursina2005} and has initiated a flurry of experimental and theoretical research activities since the discovery of its Kondo insulating properties in 2009.\cite{Strydom2009,Nishioka2009,Tanida2010b,Hanzawa2010b,Kambe2010,Khalyavin2010,
Robert2010,Lue2010,Kato2011,Matsumura2011,Kimura2011_Ru,Kimura2011_Os,Yutani2011,
Hanzawa2011,Nishioka2011,Tanida2011,Hanzawa2011a,Kondo_2011,Goraus2012} The electrical resistivity shows a thermally activated behavior at elevated temperatures and has a more metallic like form several degrees below $T_0=27$ K where a phase transition occurs as revealed by specific heat measurements.\cite{Strydom2009,Nishioka2009} A large anisotropy is observed in the static susceptibility ($\chi_a$ \textgreater $\chi_c$ \textgreater $\chi_b$) and along the easy axis the susceptibility is Curie-Weiss like above $T_0$.\cite{Nishioka2009,Tanida2010b,Yutani2011} For some time it has been a matter of debate whether the phase transition is magnetic in origin.\cite{Hanzawa2010b,Kambe2010,Khalyavin2010,Robert2010,Kato2011,Matsumura2011} In fact, in the structure of CeRu$_2$Al$_{10}$ (space group \textsl{Cmcm}) each Ce atom is situated in a polyhedral cage consisting of 16 Al and 4 Ru atoms, giving rise to very large Ce-Ce distances of more than 5\,\AA.\cite{Tursina2005} Thus, it is not obvious to explain the relative high ordering temperature using a standard model based on RKKY exchange interactions.

Nevertheless, recent muon spin relaxation and neutron diffraction experiments confirmed the magnetic nature of the phase transition. The Ce moments order antiferromagnetically along the $c$ direction with a small ordered moment of $\mu_{ord}\!\approx\!0.4$\,$\mu_B$.\cite{Kambe2010,Khalyavin2010,Robert2010,Kato2011} The latest NMR data agree with this scenario.\cite{Matsumura2011} Below $T_0$ inelastic neutron scattering (INS) experiments on a polycrystalline sample have found the existence of a spin gap of several meV, characterized by a magnetic mode with flat dispersion at about 8\,meV.\cite{Robert2010} Above $T_0$ this excitation is drastically suppressed and therefore cannot be attributed to a crystal electric field (CEF) excitation. More recent INS results have found CEF splittings of 30 and 46\,meV.\cite{AdrojaPrivCom} It has been suggested that the mixture of Kondo and CEF effects may be responsible for the small size of the ordered moment. A point charge model yields a moment of only 0.62\,$\mu_B$ along $c$ for the ground state (GS),\cite{Hanzawa2011} but the combination of large spin gap, small ordered moment and high ordering temperature remains quite puzzling. Recent optical studies have suggested an electronic structure with strong anisotropic hybridization between the $4f$ and conduction electrons which is weakest along the crystallographic $b$ direction. It has been then speculated that a charge density wave can form along $b$ which in turn may trigger the magnetic transition.\cite{Kimura2011_Ru,Kimura2011_Os}

The objective of the present work is to elucidate the local electronic structure of the $4f$ electrons. The GS wave function of the $4f$ electrons in this orthorhombic compound is expected to be highly anisotropic due to the presence of the CEF, with important consequences for the magnetic properties and the gap formation as pointed out already by several groups in their study on other orthorhombic semiconductors like CeNiSn and CeRhSb.\cite{IkedaMiyake1995,Mishchenko1997,Coleman2000,Yamada2003} 

Efforts to prove these theories experimentally turned out to be challenging because broadening of the CEF excitations due to hybridization and the existence of spin gaps prevented the determination of the ground states in these compounds by inelastic neutron scattering.\cite{KohgiCeNiSn1993,AdrojaCeNiSn1996,ParkPRB58,AdrojaCeRhSb1999,TakabatakeCeNiSn2005} Hence there is need for an alternative approach. Our method of choice is x-ray absorption spectroscopy (XAS) at the Ce M$_{4,5}$ edge which has shown to be a complementary technique to neutron scattering to determine the GS wave function of CEF split tetragonal cerium compounds.\cite{HansmannPRL100,WillersPRB80,WillersPRB81,WillersPRB85} The sensitivity to the initial state symmetry is achieved via the different absorption for light polarized $E\!\parallel\!c$ (with $c$ being the long tetragonal axis) and $E\!\perp\!c$. We now extend this method to an orthorhombic compound where the polarization dependence of the absorption, the linear dichroism (LD), has to be measured for all three directions, i.e. for $E\!\parallel\!a$, $E\!\parallel\!b$, and $E\!\parallel\!c$ in order to determine the initial state symmetry.

Single crystals of CeRu$_2$Al$_{10}$ were grown by an Al self-flux method \cite{Takabatake2010} and their quality and orientation were confirmed by Laue x-ray diffraction. All XAS measurements were carried out at the Dragon bending magnet beamline BL11A1 of the NSRRC in Taiwan. The energy resolution at the Ce M$_{4,5}$ edge ($h\nu\approx870-910$\,eV) was 0.4\,eV. The single crystals were cleaved \textsl{in situ} in an ultra high vacuum of $\sim\!10$$^{-10}$\,mbar to obtain clean sample surfaces. We measured the total electron yield (TEY) and normalized the signal to the incoming flux $I_0$ as measured on an Au-mesh at the entrance of the experimental chamber.
Two crystals were investigated: one mounted with the $c$ and the other one with the $a$ axis parallel to the Poynting vector of the incoming light. This way we were able to vary the electric field from $E\!\parallel\!a$ to $E\!\parallel\!b$ and from $E\!\parallel\!b$ to $E\!\parallel\!c$, respectively, by rotating the crystals in steps of 90$^\circ$ around the Poynting vector. The spectra for $E\!\parallel\!b$ were recorded on both samples so that the three polarizations could be normalized to each other. The data were reproduced by probing several positions on the samples and recleaving the crystals.

Ionic full multiplet calculations were performed with the XTLS 8.3 program \cite{TanakaJPSC63} to simulate the XAS spectra. Initially, the experimental isotropic spectra $I_{\rm isotropic} = I_{E \parallel a} + I_{E \parallel b} + I_{E \parallel c}$ were fitted. The best agreement was achieved with a reduction of the atomic Hartree-Fock values of about 40\% for $4f-4f$ Coulomb interactions and of about 20\% for the $3d-4f$ interactions. The reduction factors account for configuration interaction effects not included in the Hartree-Fock scheme. Their size is in agreement with previous findings on tetragonal Ce systems.\cite{HansmannPRL100,WillersPRB80,WillersPRB81,WillersPRB85} We follow the coherent approach for our simulations, i.e. the XAS spectra are calculated directly from the CEF mixed GS wave functions, the latter being fabricated via CEF parameters since for orthorhombic symmetry the XAS spectra cannot be calculated as incoherent sums of the pure $J_z$ spectra.

\begin{figure}[ht]
	\centering
	 \includegraphics[width=1.00\columnwidth]{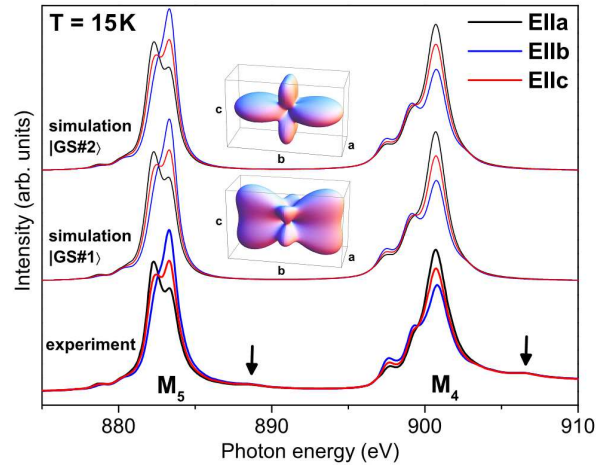}
		\caption{(Color online) Measured and simulated linear polarized XAS spectra of CeRu$_2$Al$_{10}$ at the M$_{4,5}$ absorption edge at $T=15$\,K. The two simulations reproducing the experimental data are shown with the corresponding spatial distributions of the $4f$ electrons as insets. The arrows indicate the spectral weight coming from the $4f^0$ contribution in the initial state.}
		\label{XAS_LowT}
\end{figure}

The CEF parameters are defined via the CEF Hamiltonian which arises from the expansion of the Madelung potential in spherical harmonics:
	\begin{eqnarray*}
	V(r,\theta,\Phi) = \sum^{\infty}_{k=0} \sum^k_{m=-k} A^m_k\,r^k\,C^m_k(\theta,\Phi)
	\end{eqnarray*}
$C^m_k(\theta,\Phi)$ = $\sqrt{\frac{4\pi}{2k+1}}\,Y^m_k(\theta,\Phi)$ are the semi-normalized spherical harmonics.
The expectation value $\langle r^k \rangle$ of the radial part of the wave function cannot be calculated analytically and is therefore included in the CEF parameters $\check{A}^m_k = A^m_k \langle r^k \rangle$, which are determined experimentally.

There are nine independent CEF parameters, $\check{A}^0_2$, $\check{A}^2_2$, $\check{A}^0_4$, $\check{A}^2_4$, $\check{A}^4_4$, $\check{A}^0_6$, $\check{A}^2_6$, $\check{A}^4_6$, and $\check{A}^6_6$ in the orthorhombic point group C$_{2v}$ for the cerium ion in CeRu$_2$Al$_{10}$ and the CEF splits the $J$=5/2 and 7/2 multiplets of Ce$^{3+}$ into seven Kramer's doublets. Each of these states can generally be represented in the basis of $|J,J_z \rangle$ as
	$$\sum_{J_z=-\frac{5}{2},-\frac{3}{2},...,\frac{5}{2}} \alpha_{J_z} |5/2,J_z \rangle + \sum_{J_z=-\frac{7}{2},-\frac{5}{2},...,\frac{7}{2}} \beta_{J_z} |7/2,J_z \rangle$$
for values of $J_z$ which fulfill $\sum_{J_z}(\alpha^2_{J_z} + \beta^2_{J_z})=1$ and $\Delta J_z= \pm2$.
The full multiplet routine takes the higher $J$=7/2 multiplet into account. However, for most Ce compounds the $J$=7/2 does not mix with the lower $J$=5/2 multiplet because the CEF splitting $\Delta$E$_{\textrm{CEF}}$ is much smaller than the spin orbit splitting $\Delta$E$_{\textrm{SO}}$ of $\sim\!280$\,meV. In such a case only the $\check{A}^m_k$ parameters with $k \leq 4$ affect the $J_z$ mixing of the lower $J$=5/2 multiplet and it is justified to set the higher order parameters to zero ($\check{A}^m_6$=0). For $\Delta$E$_{\textrm{SO}}$ $\gg$ $\Delta$E$_{\textrm{CEF}}$ and $\check{A}^m_6$=0 the remaining $\check{A}^m_k$ can be converted to Stevens parameters $B^m_k$.\cite{footnote}

The bottom curves in Fig.~\ref{XAS_LowT} are the measured low-temperature linear polarized XAS spectra of CeRu$_2$Al$_{10}$. The main absorption features are due to the absorption process $3d^{10}4f^1$ $\rightarrow$ $3d^94f^2$ (M$_4$ and M$_5$ edges). The spectra show a clear polarization dependence in all three crystallographic directions as expected for an orthorhombic compound. From INS experiments the splitting between the CEF ground state and the first excited CEF state is known to be $\sim\!30$\,meV,\cite{AdrojaPrivCom} so that data taken at  $T=15$\,K  are representative for the CEF ground state. There is some additional faint spectral weight at about 888 and 906\,eV in the absorption spectra (indicated by the arrows in Fig.~\ref{XAS_LowT}) due to the $3d^{10}4f^0$ $\rightarrow$ $3d^94f^1$ absorption process which is representative for the $4f^0$ contribution in the initial state. However, we note that for CeRu$_2$Al$_{10}$ this contribution is fairly small (compared with the $f^0$ in 1:2:2 compounds),\cite{WillersPRB85} which is in agreement with recent findings from $3d$ photoemission spectroscopy.\cite{Goraus2012} The small $4f^0$ contribution implies that the impact of the Kondo effect on the polarization of the $4f^1$ absorption edges is minor.

The full multiplet routine yields two different wave functions (in the following referred to as $|GS\#1\rangle$ and $|GS\#2\rangle$) that describe the experiment very well (see Fig.~\ref{XAS_LowT}). We use the $c$-axis as quantization axis and the corresponding $J_z$ coefficients are listed in Table~\ref{CF_wave_function}. Both solutions are mainly composed of the $J$=5/2 multiplet and have some small admixtures from the higher $J$=7/2 multiplet. The effect of the $J$=7/2-contributions on the LD is negligible and cannot be seen in the spectra when changing into Stevens approximation. For solution $|GS\#1\rangle$ the contribution $|5/2,\mp 3/2 \rangle$ is very strong in contrast to solution $|GS\#2\rangle$, which is dominated by $|5/2,\pm 1/2 \rangle$ and $|5/2,\pm 5/2 \rangle$. The orbitals in the insets of Fig.~\ref{XAS_LowT} represent the spatial distributions of the $4f$ electrons for the respective GS wave functions. It turns out that we can find CEF parameters for both wave functions which satisfy the results of INS, i.e. they yield the CEF transition energies $\Delta E_1=30$\,meV and $\Delta E_2=46$\,meV and the ratio of the inelastic neutron intensities $I_1/I_2=1.35$.\cite{AdrojaPrivCom} The respective CEF parameters are summarized in Table~\ref{CF_parameters}.

\begin{table}
\renewcommand{\arraystretch}{1.25}
	\begin{tabular*}{0.48\textwidth}{@{\extracolsep{\fill}} lcc}
		\hline
		$|J,J_z \rangle$				&$|GS\#1\rangle$					 &$|GS\#2\rangle$			\\
		\hline
		$|5/2,\pm 5/2 \rangle$	&$\pm 0.47(2)$ 						&$\pm 0.68(2)$				\\
 		$|5/2,\pm 1/2 \rangle$ 	&$\pm 0.32(2)$ 						&$\pm 0.73(2)$				\\
 		$|5/2,\mp 3/2 \rangle$	&$\pm 0.82(2)$ 						&$\mp 0.02(4)$				\\[2mm]
 		$|7/2,\pm 5/2 \rangle$	&$\mp 0.05(1)$						&$\mp 0.02(1)$				\\
 		$|7/2,\pm 1/2 \rangle$ 	&$\pm 0.01(1)$						&$\mp 0.03(1)$				\\
 		$|7/2,\mp 3/2 \rangle$	&$\mp 0.02(1)$						&$\pm 0.02(1)$				\\
 		$|7/2,\mp 7/2 \rangle$	&$\pm 0.03(1)$						&$\pm 0.00(1)$				\\
		\hline
	\end{tabular*}
\caption{The $J_z$ coefficients $\alpha_{J_z}$ and $\beta_{J_z}$ of the two GS wave functions which describe the XAS data. The wave functions have been calculated with the full multiplet routine, using $c$ as quantization axis. The corresponding $\check{A}_k^m$ are given in Table~\ref{CF_parameters}.}
\label{CF_wave_function}
\end{table}

\begin{table}
\renewcommand{\arraystretch}{1.25}
	\begin{tabular*}{0.48\textwidth}{@{\extracolsep{\fill}} lccccc}
		\hline
     &  $\check{A}_2^0$	&	  $\check{A}_2^2$	& $\check{A}_4^0$ & $\check{A}_4^2$	& $\check{A}_4^4$\\
		\hline
$|GS\#1\rangle$	&		4(5)			&	36(2)	&	110(5)		& -30(10) & -69(5)\\
$|GS\#2\rangle$	&		-5(2)     & 38(5) &  -40(10)  & -112(5)  &  0 (5) \\
		\hline
	\end{tabular*}
\caption{Crystal-field parameters $\check{A}_k^m$ in meV from full multiplet calculations with $\check{A}_6^m = 0$ for $m=0,\,2,\,4,\,6$. For conversion to Stevens parameters $B_k^m$ see Ref.~\onlinecite{footnote}. Note that Stevens parameters will not yield the $J$=7/2 contributions to the wave functions listed in Table~\ref{CF_wave_function}.}
\label{CF_parameters}
\end{table}

\begin{figure}[t]
	\centering
	\includegraphics[width=1.00\columnwidth]{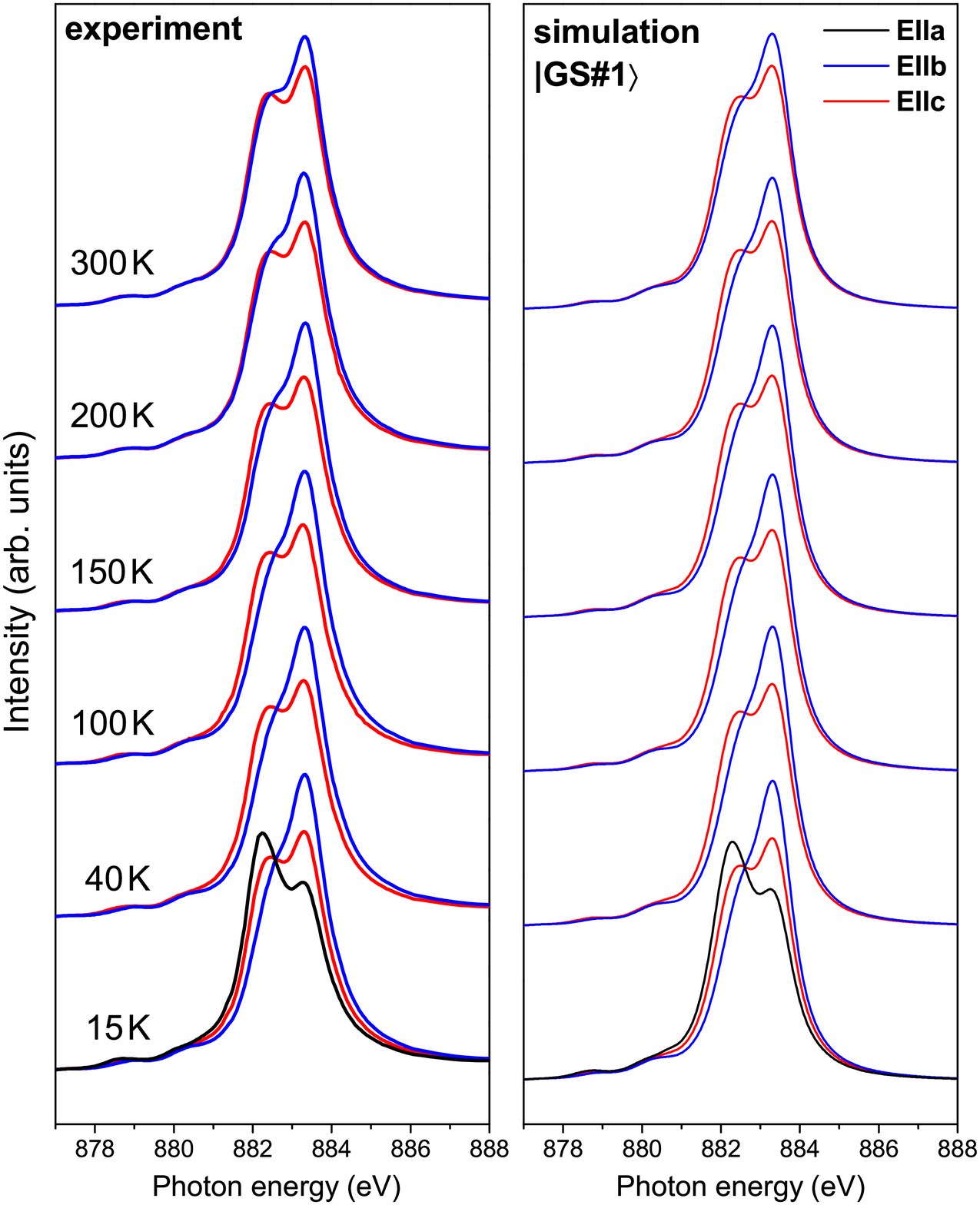}
		\caption{(Color online) Left panel: Measured temperature-dependent XAS spectra of CeRu$_2$Al$_{10}$ at the M$_{5}$ edge. Right panel: Simulation of the M$_{5}$ edge temperature dependence for $|GS\#1\rangle$. The spectra for $|GS\#2\rangle$ show the same behavior and are omitted.}
		\label{XAS_Tdep}
\end{figure}

In Fig.~\ref{XAS_Tdep} the experimental and the simulated temperature dependence of the XAS spectra are shown. Both CEF scenarios are based on the same energy splittings and at 300 K the population of the excited states at 30 and 46 meV amounts to only 20 and 10\%. Consequently both CEF models yield the same qualitative temperature dependence. We therefore omitted the simulations for $|GS\#2\rangle$ in Fig.~\ref{XAS_Tdep}. As the isotropic spectrum $I_{\rm isotropic} = I_{E \parallel a} + I_{E \parallel b} + I_{E \parallel c}$ is temperature-independent, it is sufficient to consider the temperature dependence of two polarizations and for reasons of clarity only the M$_{5}$ absorption edges for $E\!\parallel\!b$ and $E\! \parallel\!c$ are included in the illustration.  After the first cleave we measured at 15, 40, 100, 150, 200 and 300\,K, then recleaved at 300\,K and measured in the reversed temperature order. Moreover, we cross-checked and reproduced the mutual polarization for the two available samples. There is also no change of polarization up to 100 K which shows that no state gets populated, i.e. there is indeed no low lying crystal-field excitation, which is in accordance with INS results.\cite{Robert2010,AdrojaPrivCom} The change of polarization at 300 K with respect to 15 K is in agreement with excited CEF states at 30 meV and above (see right panel in Fig.~\ref{XAS_Tdep}).

\begin{figure}
	\centering
	 \includegraphics[width=1.00\columnwidth]{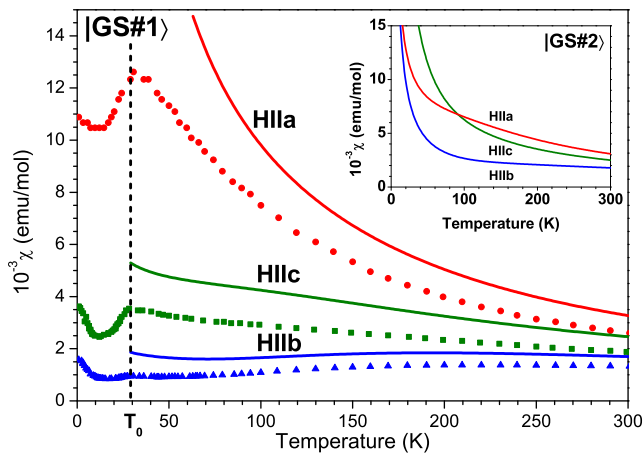}
		\caption{(Color online) Comparison of the experimental magnetic susceptibility of CeRu$_2$Al$_{10}$ as reported by Ref.~\onlinecite{Yutani2011} (symbols) and the simulated ones ($T>T_0$) from the CEF parameter sets belonging to $|GS\#1\rangle$ (lines in the main panel) and $|GS\#2\rangle$ (inset).}
		\label{Chi_both}
\end{figure}

The combined analysis of XAS and INS data still does not yield a unique solution for the ground-state wave function. We therefore calculate the temperature dependence of the static susceptibility at $H = 1$\,T in all three crystallographic directions and compare it to the experimental results found by Yutani \textsl{et al.}\cite{Yutani2011} For this purpose the full Hamiltonian, including the crystal field, the magnetic field and the spin-orbit coupling, is diagonalized and the magnetization is calculated as the Boltzmann weighted expectation value of the corresponding operator to obtain $\chi_{CEF}^\nu$ ($\nu=a,b,c$). The simulation for $|GS\#1\rangle$ is plotted in the main panel of Fig.~\ref{Chi_both} (lines) together with the data points taken from Ref.~\onlinecite{Yutani2011}. The simulation for $|GS\#2\rangle$ is included as inset. For $|GS\#1\rangle$ both the anisotropy and the qualitative temperature evolution are nicely reproduced for $T > T_0$. The $|5/2,J_z \rangle$ contributions of solution $|GS\#1\rangle$ agree quite well with the CEF analysis in Stevens approximation by Yutani et al. \cite{Yutani2011} and theoretical studies by Hanzawa who determined the $4f$ level structure of CeRu$_2$Al$_{10}$ in a point-charge model.\cite{Hanzawa2011} Note that Hanzawa uses the $b$-axis as quantization axis. The change of coordinate system from $abc \parallel yzx$ to $abc \parallel xyz$ can be carried out according to Ref.~\onlinecite{McPhaseManual}.
In contrast, solution $|GS\#2\rangle$ does not match the measured susceptibility well. It yields a cross-over of $\chi_{CEF}^a$ and $\chi_{CEF}^c$ at about 100\,K so that we exclude it as a possible ground-state wave function.

The CEF ground-state wave function $|GS\#1\rangle$ yields the magnetic moments $\mu_{CF}^{a,b,c}$ = (1.44,0.19,0.38)\,$\mu_B$. Here an infinitesimal small temperature and magnetic field along the respective directions $a$, $b$, or $c$ have been taken into account. These moments agree very well with values from magnetization at 55\,T and 1.3\,K, which find 1.3 and 0.3\,$\mu_B$ along the easy axis $a$ and the ordering axis $c$, respectively.\cite{Kondo_2011} Also, the small moment of $\mu_{CF}^{c}$~=~0.38\,$\mu_B$ agrees well with the experimental findings for the ordered magnetic moment,\cite{Kambe2010,Khalyavin2010,Robert2010,Kato2011} indicating that CEF effects are responsible for the small size of $\mu_{ord}$.

The remaining discrepancy between the measured and CEF-only susceptibility $|GS\#1\rangle$ in Fig.~\ref{Chi_both} can be explained by considering corrections due to molecular and/or exchange fields and anisotropic hybridization between $4f$ and conduction electrons. Such an anisotropic hybridization has been observed in optical conductivity measurements \cite{Kimura2011_Ru} and the resulting anisotropic Kondo interactions will affect the anisotropy of the static susceptibility. 

In conclusion, the full multiplet simulation yields two GS wave functions which reproduce the low-temperature XAS data very well. By combining our XAS with susceptibility results \cite{Yutani2011} we unambiguously identify $|GS\#1\rangle$ (see Table~\ref{CF_wave_function})
as the GS wave function for CeRu$_2$Al$_{10}$, which is in good accordance with the theoretical findings by Hanzawa.\cite{Hanzawa2011} We can further give a set of CEF parameters that satisfies XAS, susceptibility, and INS data,\cite{AdrojaPrivCom} and the resulting moments are in accordance with moments from magnetization measurements.\cite{Kondo_2011} The fact that the CEF-only moment along $c$ amounts to only $\mu^c_{CF}$\,=\,0.38\,$\mu_B$ implies that the small value of the ordered moment \cite{Kambe2010,Khalyavin2010,Robert2010,Kato2011} can be explained solely with CEF effects. The absence of any temperature dependence in the XAS data up to 100\,K confirms that the excitation at 8\,meV in the inelastic neutron spectra \cite{Robert2010} is not a CEF transition. The above results show that the selection rules for linear polarized light make soft XAS a powerful tool to determine the CEF ground state wave function of orthorhombic Kondo insulators even in the presence of spin gaps.

This work was supported by DFG grant AOBJ 583872, Germany and KAKENHI No. 20102004 of MEXT, Japan.


\begin{thebibliography}{38}
\bibitem[{foo()}]{footnote}
\bibinfo{note}{The relationship between the crystal-field parameters in
  spherical harmonics and Stevens approximation for $f^1(J\!=\!\frac{5}{2})$
  is: $\check{A}^0_2 \leftrightarrow -35 B^0_2$, $\check{A}^{-2}_2
  \leftrightarrow -\frac{7}{2} \sqrt{\frac{50}{3}} B^2_2$, $\check{A}^2_2
  \leftrightarrow -\frac{7}{2} \sqrt{\frac{50}{3}} B^2_2$, $\check{A}^0_4
  \leftrightarrow 1260 B^0_4$, $\check{A}^2_4 \leftrightarrow 63 \sqrt{10}
  B^2_4$, $\check{A}^{-2}_4 \leftrightarrow 63 \sqrt{10} B^2_4$, $\check{A}^4_4
  \leftrightarrow 18 \sqrt{70} B^4_4$, $\check{A}^{-4}_4 \leftrightarrow 18
  \sqrt{70} B^4_4$}.
\end{thebibliography}
\end{document}